\documentstyle[12pt,prl,aps]{revtex}

\begin{document}
\title{Polaronic Excitons in $\rm \bf Zn_{x}Cd_{1-x}Se/ZnSe$ Quantum Wells}
\author{Andrea De Nardis, Vittorio Pellegrini, Raffaele Colombelli and Fabio Beltram}
\address{Scuola Normale Superiore and INFM, I-56126 Pisa, Italy}
\author{I.N.Krivorotov and K.K.Bajaj}
\address{Emory University, Atlanta, GA 30322}
\author{\bf{Submitted to Phys. Rev. B}}
\maketitle
\begin{abstract}
 
We present a detailed investigation of excitonic absorption 
in $Zn_{0.69}Cd_{0.31}Se/ZnSe$ quantum wells under 
the application of a perpendicular magnetic field. 
The large energy separation between heavy- and light-hole
excitons allows us to clearly resolve and 
identify magneto-excitonic absorption resonant with the 
continuum edge of the 1S heavy-hole exciton.
Experimental values of the exciton binding energy
are compared with results of a theoretical model that
includes the exciton-phonon interaction.
The remarkable agreemeent found unambiguously indicates the 
predominant polaronic
character of excitons in $ZnSe$-based heterostructures.
 
\end{abstract}  
 
\bigskip
In recent years wide-gap II-VI semiconductor based heterostructures have
attracted much attention mostly in light of their potential
for the development of opto-electronic devices operating in 
the blue-green spectral region\cite{laser1}. 
Owing to their relatively large exciton 
binding energies ($E_{b}$), they also offer
the possibility of realizing quantum wells (QWs)
in which excitonic recombination is dominant even at 
room temperature\cite{laser2}.
These systems also present other unique properties such as
a strong exciton-phonon interaction which makes them ideal
candidates for the investigation of these kind of interactions (from
here on referred to \emph{polaronic effects}) on the
optical properties.
\par
Polaronic effects have been extensively
studied in the past\cite{polaron1,matsuura},
and significant polaron-related modifications of
fundamental optical properties were demonstrated in bulk ionic semiconductors.
Much less attention, on the contrary,
was paid to this issue in systems with reduced dimensionality, e.g. QWs.
In fact in the numerous experimental analyses on excitonic effects
in II-VI QWs \cite{beznse1,moznse1,moznse2}, the role of polarons in 
determining the exciton binding
energy was not adequately addressed, and was actually often overlooked.
This was probably due to the
large body of studies focused on GaAs-based
heterostructures\cite{mogas1,mogas2} where polaronic effects are
indeed negligible owing to the low-ionicity of the atomic bonds.
\par
In this communication we address this issue by
analyzing the magneto-absorption properties of 
$Zn_{x}Cd_{1-x}Se/ZnSe$ QWs. Our results clearly demonstrate
the important role played by polaronic effects in 
determining excitonic optical properties.
We shall show that by \emph{including polaronic effects} 
using an effective potential first derived by \emph{Aldhrich and
Bajaj}\cite{ABPot} within an envelope-function formalism it is possible to
reproduce quantitatively our experimental results 
with no adjustable material parameters.
\par
Samples studied were grown by solid source molecular-beam epitaxy 
on $GaAs(001)$
substrates. They consist of a 0.5-$\mu$m $GaAs$ buffer layer grown at
580 $^{o}$C followed by 1.5-$\mu$m-thick $ZnSe$ layer grown at 290$^{o}$ C.
Ten $Zn_{x}Cd_{1-x}Se/ZnSe$ QW's were then 
grown at 250$^{o}$ C with a 30-s interruption at each interface.
A 500-nm-thick cap layer concluded the growth.
$ZnSe$ barrier width is 30 nm, 2-, 3-, 4- and 5-nm-thick
QWs were examined. 
In order to detect transmission signals,
circular regions of about $6 \times 10^{-4} cm^{-2}$ were 
selectively removed using
standard photolithographic and wet-etching techniques.
Further details about these structures
can be found in Ref. \cite{sam}.
The separation between the energy
positions of 1S heavy- and light-hole excitons caused by
the compressive strain in the QWs, allowed us to resolve the
absorption associated to the continuum edges, even at zero magnetic field.
\par
It should be noted that despite the large Cd content the 
1S heavy-hole exciton absorption peak at T=1.6K displays a full width at
half maximum of about 9 meV indicating
good sample quality. This was also confirmed by the small
Stokes shift of the photoluminescence signal ($\approx$1 meV).
Samples were mounted on a variable-temperature insert 
and magnetic fields $B$ between 0 and 10T were applied  parallel to the
growth direction.
Magneto-absorption was studied using a 100W xenon lamp as the source.
Light was focused and collected along the growth direction
by optical fibers. Transmitted 
intensity was normalized to the incident light
in order to detect absorption changes.
\par
Figure\,1 shows typical spectra at T=1.6K 
for the sample with QW thickness $L_{w} = 5$ nm 
without (dashed line) and with an applied magnetic field (10 T, solid line).
The two peaks at 2528 meV and 2610 meV observed in the B=0 spectrum
are associated to
the 1S (hh1-e1) heavy- and (lh1-e1) light-hole exciton, respectively.
The absorption of what is supposed to be 
the heavy-hole continuum edge is observed at around 2560 meV.
Thanks to the high confinement potential, an 
additional higher-energy peak due to the 1S (hh2-e2) heavy-hole
exciton\cite{JCG} is also observed.
\par
Upon application of the magnetic field, the 1S heavy- and light-hole 
exciton peaks display a small blue-shift and no appreciable oscillator
strength enhancement. This behavior is immediately understood in
light of their
large binding energy in comparison to the cyclotron energy ($E_{c}=$5.2 meV
at 10 T). 
The light-hole exciton, however, displays a peculiar 
non-monotonic behavior which may be linked to valence-band mixing
effects \cite{moznse1}.
The heavy-hole continuum edge, on the contrary, is dramatically
altered, and a sharp peak on its low-energy
side gradually emerges with increasing magnetic fields 
(see the right panel of Fig.~1 where
absorption spectra in the region of the continuum edge
are displayed at three different magnetic fields).
This feature can be unambiguously associated to the 
heavy-hole 2S excitonic level because of
its diamagnetic behavior in the range $6T \le B \le 10$ T (see
Fig. 2).
The diamagnetic shift of that level  ($\delta
E \approx 4 meV$) indicates that the Coulomb interaction  is still
significant in the investigated range of 
magnetic fields.
A further peak, visible from 8.5 T is identified with 
the 3S excitonic level. In this magnetic field range, however, this 
exciton state acquires a Landau-like magnetic-field dependence
that, contrary to the 2S exciton case, 
indicates the predominance of the free-carrier cyclotron energy $E_{c}$ 
over the exciton binding energy.
This is in qualitatively agreement with theory developed in Ref. \cite{th}
\par
Figure 2 shows the peak energy positions of heavy- and light-hole
excitonic states (solid circles) as a function of B
for the sample with $L_{w}=5 $nm (a) and  $L_{w}=4 $ nm (b).
>From the expected values of 1S and 2S position to B=0 we can determine
the energy difference bitween the binding energis of 1S and 2S states,
adding the calculated values of $E^{2S}_{b}$\ we can determine the
binding energy of the 1S state of the $hh$ exciton.
\par
Our theoretical analysis based on variational calculation 
of 1S exciton binding energy versus quantum well width 
demonstrates the importance of polaronic effects in 
$Zn_{x}Cd_{1-x}Se/ZnSe$ QW's. A complete account of 
the polaronic effects leads to an effective Hamiltonian 
for the exciton in which not only electron and hole band masses 
are replaced by the corresponding electron and hole polaron effective
masses but also the effective potential differs from the
Coulomb potential screened by the static dielectric constant 
$\epsilon_{0}$ \cite{polaron1}. 
\par
The Hamiltonian of the exciton (apart from the electron and hole
self-energy terms) 
can be expressed as a sum of three terms:
\begin{equation}
H=H_{1}+H_{2}+H_{ex}
\end{equation}
where $H_{1}$ and $H_{2}$ describe the motion of electron polaron 
and hole polaron along the growth axis (z):
\begin{equation}
H_i=-\frac{\hbar^2}{2m^*_i}\frac{\partial^2}{\partial z^2}+V_i(z_i)
\end{equation}
where $V_1(z_1)$ and $V_2(z_2)$ are the confining potentials for the electron 
polaron and the hole polaron respectively, and $m^*_1$ and $m^*_2$ are
the electron and hole polaron masses along the z axis.
\par
 The term $H_{ex}$ describes the internal motion of the exciton in the
x-y plane:
\begin{equation}
H_{ex}=-\frac{\hbar^2}{2\mu^*}\left(\frac{1}{\rho}
\frac{\partial}{\partial\rho}\left(\rho\frac{\partial}{\partial\rho}\right)
+\frac{1}{\rho^2}\frac{\partial^2}{\partial\phi^2}\right)+V_{int}(r)
\end{equation}
where $\mu^*$ is the reduced electron and hole polaron mass in the x-y plane,
$\rho$ and $\phi$ are polar coordinates in the x-y plane and 
$r^2=(z_1-z_2)^2+\rho^2$.
Hole masses are expressed in terms of the Luttinger parameter in these
formulas:
\par
For the effective interaction potential $V_{int}$, we use the following
two different forms. First we consider the Coulomb potential screened
by the static dielectric constant i.e.
\begin{equation}
V_C=-\frac{e^2}{\epsilon_0 r}
\end{equation} 
and second, we consider the potential $V_{AB}$ derived by 
Aldrich and Bajaj \cite{polaron1}  that 
takes into account the polaronic effects.
\begin{eqnarray}
V_{AB}=-\frac{e^{2}}{\epsilon_{0}r}
-\frac{e^{2}}{2\epsilon^\prime r}(exp(-\beta_{1}r)+exp(-\beta_ir))
\nonumber \\
+\sum_{i=1}^2\frac{e\beta_i}{2\epsilon^{\prime}}\frac{exp(-\beta_ir)}
{1+\alpha_i/12+\alpha_i/(4+\alpha_i/3)}
\end{eqnarray}
where:
\begin{equation}
\beta_i=(\frac{2m_i\omega}{\hbar})^{1/2}
\end{equation}
\begin{equation}
\alpha_i=\frac{e^2\beta_i}{2 \epsilon^\prime\hbar \omega}
\end{equation}
\begin{equation}
\frac{1}{\epsilon^{\prime}}=\frac{1}{\epsilon_{\infty}}-\frac{1}{\epsilon_{0}}
\end{equation}
\begin{equation}
m_\alpha = m_0 \frac{1-\alpha / 12}{1+ \alpha / 12}
\end{equation}
where $\omega$ is the longitudinal optical phonon frequency, 
$\epsilon_\infty$ is high frequency dielectric constant, and
$m_i$ is electron (hole) band mass. 
\par
In order to calculate the binding energy of 1S state of an exciton as 
a function of well width we follow a variational approach and
use the following form of the trial wave function:
\begin{equation}
\Psi(z_{1},z_{2},\rho,\phi)=F_{1}(z_{1})F_{2}(z_{2})
exp(-\lambda r-\nu \rho^{2}-\eta (z_{1}-z_{2})^{2})
\end{equation}
where $F_1(z_1)$ and $F_2(z_2)$ are ground state eigenfunctions of 
the electron polaron and hole polaron Hamiltonians $H_{1}$ and $H_{2}$ 
respectively.
\par
The binding energy $E_{B}$ of the 1S state of the exciton is then defined as:
\begin{equation}
E_B=E_1+E_2-min_{\lambda,\nu,\eta}(\frac{<\Psi|H|\Psi>}{<\Psi|\Psi>})
\end{equation}
where $E_{1}$ and $E_{2}$ are ground state eigenvalues of 
the electron polaron and hole polaron Hamiltonians $H_{1}$ and $H_{2}$ 
respectively.

\par
In our calculation we have used the following values of various physical 
parameters for the  $Zn_{x}Cd_{1-x}Se/ZnSe$ QW's. Electron band
offset $\Delta E_e = 0.230$ eV, hole band offset $\Delta E_h = 0.115$ eV,
electron polaron mass for the  barrier material $m_1^b=0.155 m_0$,
electron polaron mass for the well  material $m_1^w=0.14 m_0$,
static dielectric constant $\epsilon_0 = 8.7$, high frequency
dielectric constant $\epsilon_\infty = 5.73$, LO phonon energy
$E_{LO}=0.0317 eV$, Luttinger parameters $\gamma_1 =2.45$ 
and $\gamma_2 =0.61$.

The results of the calculation of binding energy versus well width for 
both $V_C$ and $V_{AB}$ are shown in Figure 3. The use of simple 
screened Coulomb potential leads to essential underestimation of 
the exciton binding energy in $Zn_xCd_{1-x}Se/ZnSe$ QW's. 
On the other hand, the values of binding energy obtaind with $V_{AB}$
are in a good agreement with the experimentally measured binding
energies for the samples with $L_w=5$ nm and  $L_w=4$ nm.
 
\par
In conclusion, we have measured ... 
Using variational approach, we have calculated the exciton binding energies 
for $Zn_{x}Cd_{1-x}Se/ZnSe$ QW's vs. quantum well width using both Coulomb 
potential screened by static dielectric constant $\epsilon_{0}$ and the 
potential $V_{AB}$ derived by Aldrich and Bajaj, that takes account of 
polaronic effects. The comparision of the calculated values of the exciton binding energy to the experimantally obtained binding energies for 4 nm and 5 nm 
quantum wells demonstrates that the potential $V_{AB}$ should be used in 
the calculations of binding energy in order to achieve a reasonable agreement 
with the experimantal data and that the polaronic effects essentially influence
physical properties of excitons in $ZnSe$-based heterostructures.
Detailed investigation of the diamagnetic shift will be presented elsewhere.

\begin{figure}
\caption{Left Panel: Absorption spectra for the
$Zn_{0.69}Cd_{0.31}Se/ZnSe$ sample with 5 nm well width
at magnetic fields B=0T and B=10T.
The magnetic fields are applied along the growth directin 
(Faraday configuration). Right Panel: Absorption spectra 
of the heavy-hole continuum edge at three different magnetic fields.}
\end{figure}
 
\begin{figure}
\caption{(a) and (b) Energy of exciton absorption peaks vs. magnetic field for
both the two sample (circles).
Orizontal arrows indicate continuum energy position calculated
by the polaron-based model. Least-square fit 1s and 2s hh1-e1 are shown as
dashed curves.}
\end{figure}

\begin{figure}
\caption{Calculated values of exciton binding energy vs. well
width for the $Zn_{0.69}Cd_{0.31}Se/ZnSe$ quantum well. 
Solid line represent exciton binding energy as obtained using Coulomb
potential $V_{C}$. Dashed line represent exciton binding energy as 
obtained using potential $V_{AB}$. Solid squares represent experimentally 
measured binding energies for 4 nm and 5 nm quantum wells. }
\end{figure}
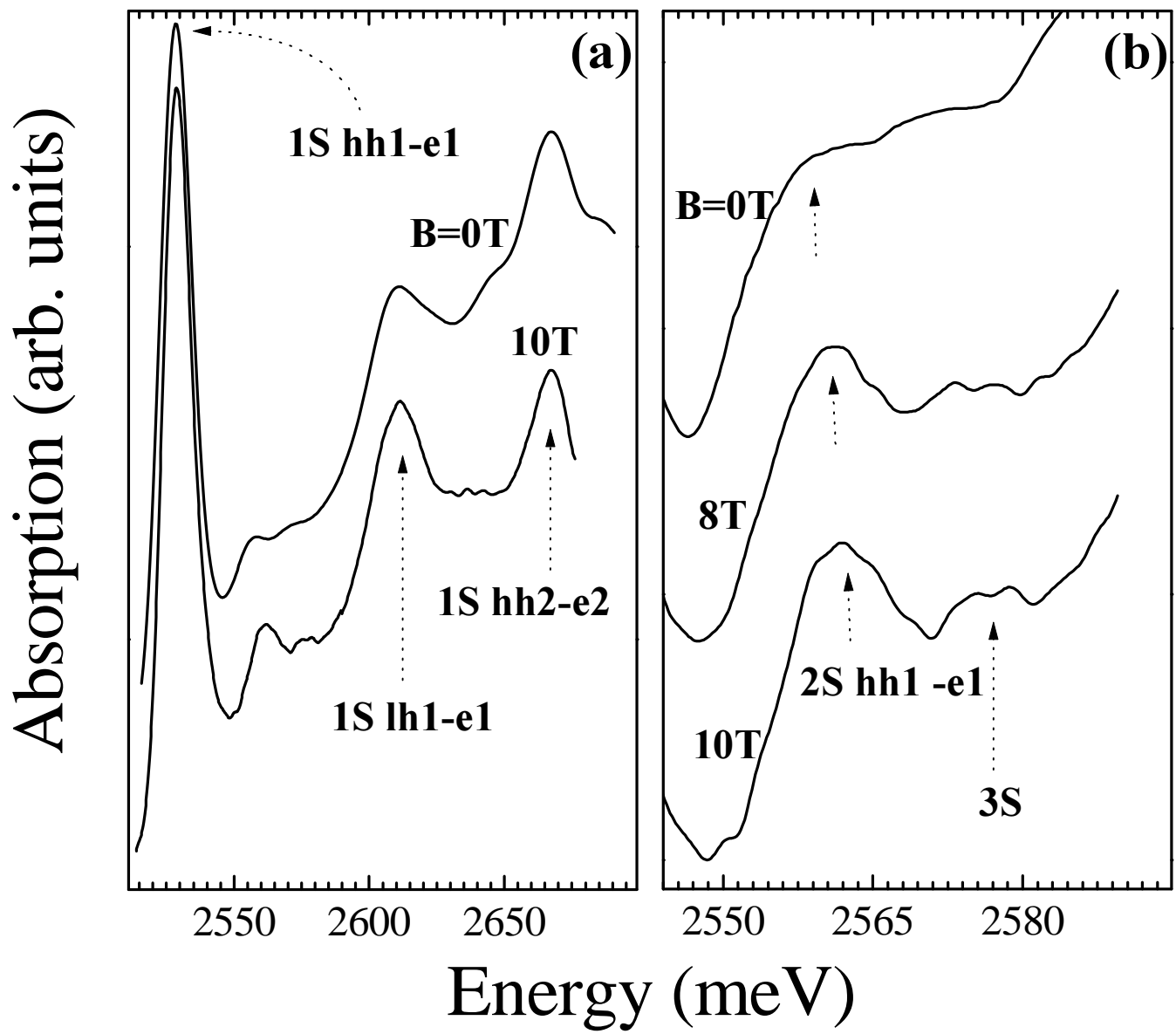
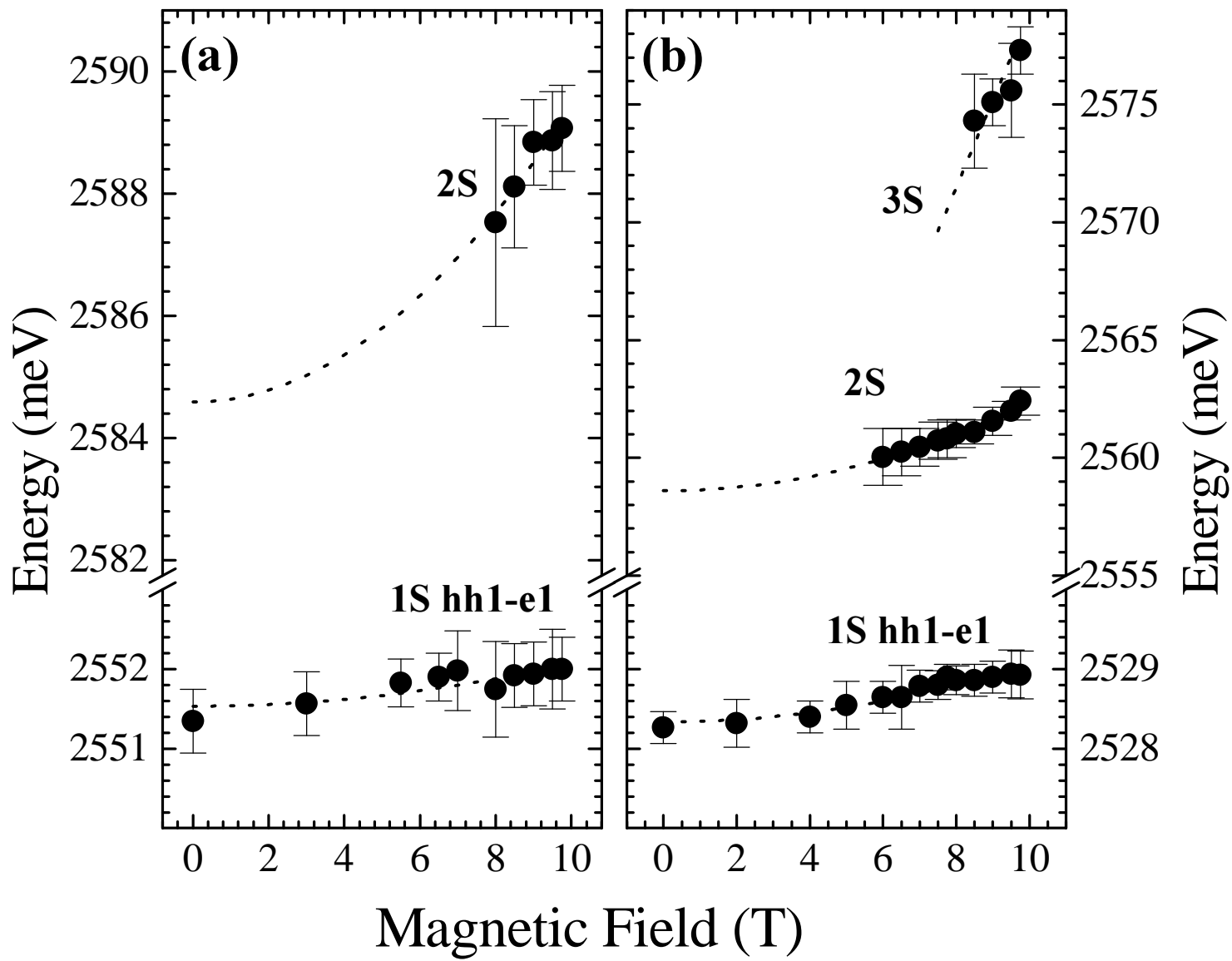
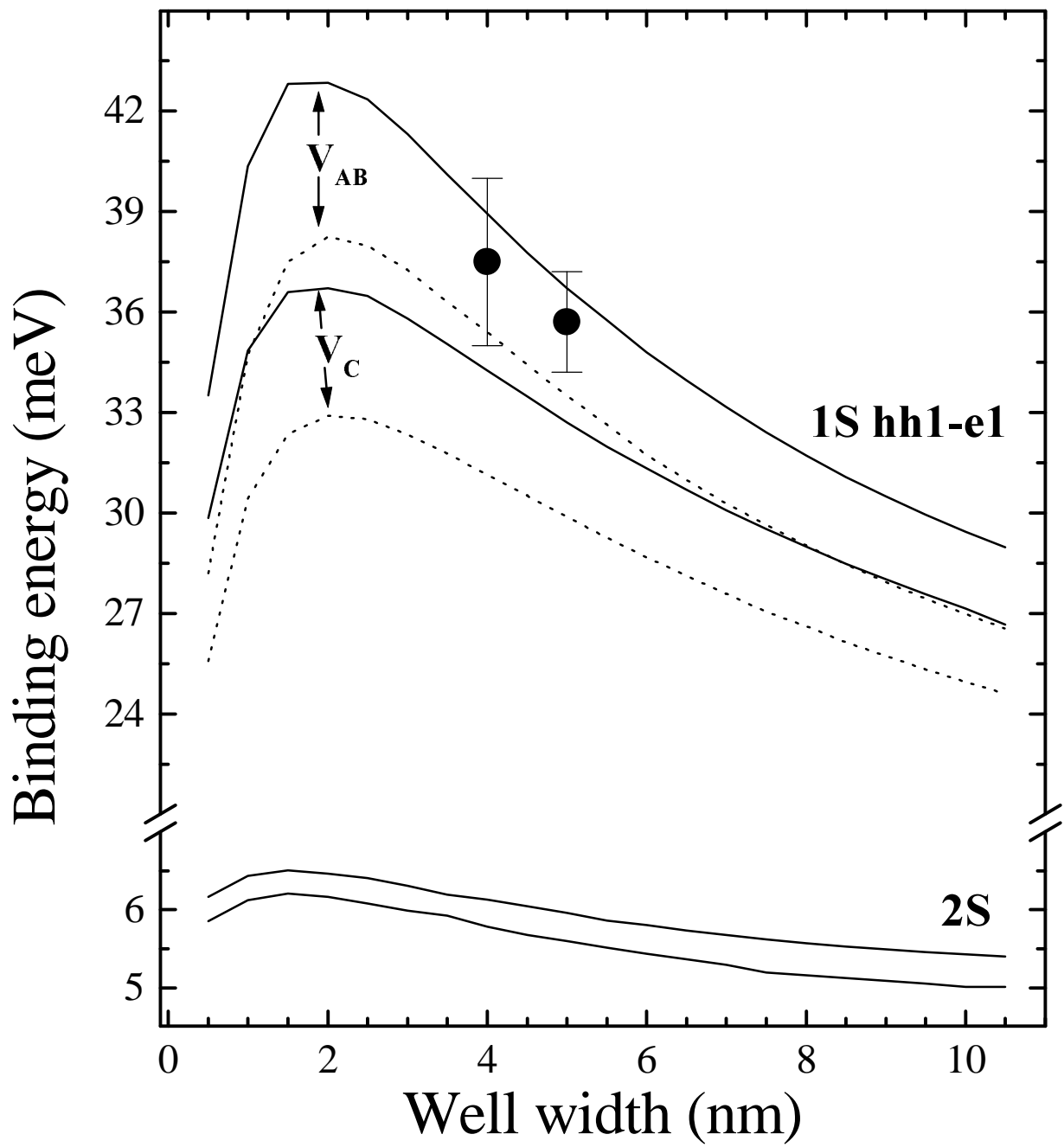
\end{document}